\documentclass[%
 reprint,
 amsmath,amssymb,
 aps,
]{revtex4-2}

\usepackage{graphicx}
\usepackage{dcolumn}
\usepackage{bm}
\usepackage{braket} 
\usepackage{xcolor}
\begin{document}

\preprint{APS/123-QED}

\title{Testing Supersymmetric Hidden Sectors with Long-Baseline Atom Interferometers}

\author{Oem Trivedi$^{}$}
\affiliation{$^{}$Department of Physics and Astronomy, Vanderbilt University, Nashville, TN, 37235, USA}

\email{Email: oem.trivedi@vanderbilt.edu}

\date{\today}

\begin{abstract}
Atomic interferometry provides a sensitive near Earth probe of high energy physics through precision measurements of quantum phase. In this Letter, we point out that MAGIS and AION like long-baseline atom interferometers can also be used to test supersymmetric hidden sectors, if these sectors contain ultralight moduli, dilatons or hidden scalars that induce coherent phase oscillations. In such a setup, the measured atomic phase does not only constrain an effective phenomenological scalar coupling. It can be related to derivatives of supersymmetric gauge kinetic functions, Kähler metrics, Yukawa couplings, Higgs sector parameters and the QCD scale along light hidden sector directions. We derive the mapping from a generic SUSY/SUGRA modulus to the effective atom interferometric coupling, and show that future phase sensitivities may probe very small visible sector admixtures of otherwise hidden fields. This identifies MAGIS/AION type experiments as non-collider probes of supersymmetric and string-motivated infrared relics, complementary to gravitational wave, astrophysical and collider searches.

.

\end{abstract}

\maketitle

Atomic interferometry has seen a lot of interest in high energy and gravitational physics in recent years, mainly because it gives a very clean way of combining quantum coherence, inertial sensing and long-baseline measurements in a single experimental setup \cite{at1Cronin:2009zz,at2Abend:2020djo,at3Buchmueller:2023nll}. The basic idea is rather simple: freely falling atoms can be treated as quantum test masses, and the phase of their matter waves responds to accelerations, tidal fields, gravitational radiation and also to possible time-dependent variations of microscopic energy scales. This has opened a useful new window for fundamental physics, where laboratory or near-Earth instruments can be sensitive to ultralight dark matter, equivalence principle violation, fifth forces and related effects \cite{at4Abdalla:2024sst,at5TVLBAIProto:2025wyn,at6Roura:2024rmm,at7Werner:2024yxw,at8Premawardhana:2023hdz}. Unlike conventional particle searches, these experiments do not need to produce the new particle directly. Instead, they look for the coherent phase shift induced by weakly coupled fields or by spacetime perturbations. This is why atom interferometers are particularly interesting for sectors which are far below collider thresholds, but may still have large coherent occupation numbers or appreciable gravitational imprints.
\\
\\
A major part of this interest has come from the connection of atom interferometers with gravitational wave detection \cite{gw1Dimopoulos:2007cj,gw2Graham:2012sy,gw3enhGraham:2016plp,gw4AEDGE:2019nxb,gw5Canuel:2019abg,gw6Geiger:2015tma,gw7impArvanitaki:2016fyj}. This has motivated proposals such as the Matter wave Atomic Gradiometer Interferometric Sensor (MAGIS) \cite{magis1abe2021matter,magis2adamson2018proposal,magis3Coleman:2018ozp,magis4Mitchell:2022zbp,magis5Jachinowski:2021jcm} and the Atom Interferometer Observatory and Network (AION) \cite{aion1Badurina:2019hst,aion2buchmuller2018atom,aionsenseBadurina:2021lwr,aionsense2Badurina:2022ngn,aionsense3Badurina:2023wpk,aionsense4Badurina:2021rgt}. In these experiments, separated atomic interferometers are used as differential phase sensors over long baselines. One of the main motivations behind MAGIS and AION is to probe the mid-frequency gravitational wave band, which lies between the most sensitive regimes of terrestrial laser interferometers and space-based interferometers. At the same time, the same instruments can also search for ultralight dark matter and other very weakly coupled sectors.
\\
\\
In practice, atomic wave packets are split, redirected and recombined using laser pulses, and the resulting phase differences are compared between atomic ensembles at different spatial locations. The gradiometric arrangement is important because it suppresses common laser noise, while still keeping sensitivity to tidal gravitational signals, coherent oscillations of fundamental constants and long range forces. Thus the staged versions of MAGIS and AION are not only pathfinders for atom interferometric gravitational wave detection, but also precision probes of new physics in the sub-Hz to Hz range. In this letter, we point out that the same kind of setup can also be used to test supersymmetric hidden sectors in a way which is rather different from collider searches or conventional astrophysical probes.
\\
\\
To see how this works, let us first recall the basic response of a light pulse atom interferometer. Laser pulses split, redirect and then recombine an atomic matter wave. For a simple acceleration-sensitive configuration, the leading phase shift is
\begin{equation}
\Delta \phi_{\rm AI}
\simeq
k_{\rm eff}aT^2
\end{equation}
where $k_{\rm eff}$ is the effective wave vector, $a$ is the acceleration along the interferometer axis, and $T$ is the pulse separation time. Therefore, unlike LIGO, which measures mirror displacement \cite{LIGO1Scientific:2016aoc,LIGO2Scientific:2007fwp}, MAGIS and AION measure quantum phase shifts accumulated by freely falling atoms. In a long-baseline configuration, two atomic ensembles at positions $z_1$ and $z_2$ are interrogated by common laser pulses. The observable is then the differential, or gradiometer, phase
\begin{equation}
\Delta \Phi_{\rm grad}(t)=
\Delta \phi(z_2,t)-\Delta \phi(z_1,t)
\simeq
k_{\rm eff}T^2
\left[
a(z_2,t)-a(z_1,t)
\right]
\end{equation}
For an acceleration field that varies slowly across the baseline $L=z_2-z_1$, this gives
\begin{equation}
\Delta \Phi_{\rm grad}(t)
\simeq
k_{\rm eff}T^2L
\frac{\partial a}{\partial z}
\end{equation}
so the same instrument is directly sensitive to tidal fields, gravity gradients and propagating metric perturbations, while the common mode laser noise is strongly suppressed.
\\
\\
For gravitational waves, the strain $h(t)$ changes the light propagation time, and therefore also the phase imprinted on the separated atomic ensembles. The response may be written schematically as
\begin{equation}
\Delta \Phi_{\rm GW}(f)=
{\cal R}_{\rm GW}(f)h(f)
\end{equation}
with the leading scaling
\begin{equation}
\Delta \Phi_{\rm GW}
\sim
k_{\rm eff}Lh
\end{equation}
and, for resonant pulse sequences with enhancement factor $Q$ \cite{gw3enhGraham:2016plp},
\begin{equation}
\Delta \Phi_{\rm GW}
\sim
Qk_{\rm eff}Lh
\end{equation}
Thus, a longer baseline, large momentum transfer atom optics, and resonant enhancement all improve the strain response. For representative values $k_{\rm eff}\sim10^{10}\,{\rm m}^{-1}$, $L\sim100 \,{\rm m}$, and $h\sim10^{-20}$ \cite{at3Buchmueller:2023nll}, one finds
\begin{equation}
\Delta \Phi_{\rm GW}
\sim
10^{-8} {\rm rad}
\end{equation}
This is a small signal, but it is coherent, and importantly it is a phase signal rather than a displacement signal.
\\

The same experimental architecture is also sensitive to coherent scalar dark matter \cite{aionsenseBadurina:2021lwr,aionsense2Badurina:2022ngn,aionsense3Badurina:2023wpk,aionsense4Badurina:2021rgt,gw7impArvanitaki:2016fyj}. A light scalar making up a fraction $F_{\rm DM}$ of the local dark matter density behaves, on laboratory timescales, as
\begin{equation}
\varphi(t)= \varphi_0\cos(m_\varphi t+\beta),
\qquad
\varphi_0=
\frac{\sqrt{2F_{\rm DM}\rho_{\rm DM}}}{m_\varphi}
\end{equation}
and this gives a narrow spectral feature at
\begin{equation}
f_\varphi=
\frac{m_\varphi}{2\pi\hbar}
\end{equation}
For $m_\varphi\sim10^{-15} \,{\rm eV}$, this lies in the sub-Hz/Hz range which is relevant for long baseline atom interferometers. If the scalar produces a small oscillation in an atomic transition frequency,
\begin{equation}
\omega_A(t)=
\omega_A+\delta\omega_A(t)
\end{equation}
then the corresponding phase shift is
\begin{equation}
\Delta \Phi_\varphi=
\int dt,g(t)\delta\omega_A(t)
\sim
\omega_A T_{\rm eff}
\frac{\delta\omega_A}{\omega_A}
\end{equation}
where $g(t)$ is the sensitivity function and $T_{\rm eff}$ is the effective interrogation time. The scalar dark matter part of the signal therefore appears as an oscillatory phase contribution,
\begin{equation}
\Delta \Phi_\varphi(t)= A_\varphi\cos(m_\varphi t+\beta)
\end{equation}
whose amplitude is set by the fractional modulation of the atomic transition frequency. The full measured phase time series may then be written as
\begin{equation} \label{eq:phase_decomposition}
\Delta \Phi(t) = 
\Delta \Phi_{\rm GW}(t)
+
\Delta \Phi_\varphi(t)
+
\Delta \Phi_{\rm grad}(t)
+
\Delta \Phi_{\rm noise}(t)
\end{equation}
Here the gravitational wave term is proportional to the strain through ${\cal R}_{\rm GW}(f)$, the scalar term is a coherent oscillation of atomic energy scales, the gradiometer term accounts for ordinary tidal and environmental gradients, and the last term denotes instrumental and environmental noise. The task experimentally is then to look for coherent, correlated and frequency-localized phase signals, and to use their dependence on baseline, pulse sequence, atomic species and detector geometry to separate them from backgrounds.
\\
\\
Starting from Eq.~(\ref{eq:phase_decomposition}), the relevant point for hidden sector physics is that any field capable of producing a coherent modulation of atomic transition energies contributes directly to $\Delta \Phi_\varphi(t)$. For a light scalar field $\varphi$ constituting a fraction $F_{\rm DM}$ of the local dark matter density, the field may be treated over laboratory timescales as a classical coherent oscillator
\begin{equation}
\varphi(t)
=
\varphi_0 \cos(m_\varphi t+\beta)
\end{equation}
with amplitude given by
\begin{equation}
\varphi_0
=
\frac{\sqrt{2F_{\rm DM}\rho_{\rm DM}}}{m_\varphi}
\end{equation}
If the scalar induces a fractional modulation of an atomic transition frequency, the interferometric contribution can be written as
\begin{equation}
\Delta \Phi_\varphi
\simeq
\omega_A T_{\rm eff}
\frac{\delta \omega_A}{\omega_A}
\end{equation}
where $\omega_A$ is the atomic transition frequency and $T_{\rm eff}$ is the effective interrogation time determined by the pulse sequence. The central observable is therefore an oscillatory atomic phase given as
\begin{equation}
\Delta \Phi_\varphi(t)
=
A_\varphi
\cos(m_\varphi t+\beta)
\end{equation}
with amplitude
\begin{equation}
A_\varphi
=
\omega_A T_{\rm eff}
d_{\rm eff}
\kappa
\frac{\sqrt{2F_{\rm DM}\rho_{\rm DM}}}{m_\varphi}
\end{equation}
where $\kappa=\sqrt{4\pi G_N}$ and $d_{\rm eff}$ is the effective scalar coupling to the atomic transition being probed. Thus, a null search for an oscillatory phase signal at frequency $m_\varphi/2\pi\hbar$ implies the symbolic constraint
\begin{equation}
d_{\rm eff}
<
\frac{A_{\rm lim}}{\omega_A T_{\rm eff}}
\frac{m_\varphi}{\kappa\sqrt{2F_{\rm DM}\rho_{\rm DM}}}
\end{equation}
Please note that this expression should be understood as the general experimental constraint, the actual bound depends on the instrument specific response function, the atomic species, the pulse sequence, the observation time, the final measured value of $A_{\rm lim}$ etc. In particular, the replacement
\begin{equation}
A_{\rm lim}
\rightarrow
A_{\rm lim}(m_\varphi;L,k_{\rm eff},T_{\rm eff},Q,S_\Phi,{\cal R}_\varphi)
\end{equation}
must be made in a complete experimental analysis.
\\

The connection to supersymmetric hidden sectors follows by identifying $\varphi$ with a light modulus, dilaton or hidden sector scalar belonging to a chiral superfield,
\begin{equation}
{\cal M} = M+\sqrt{2}\theta\psi_M+\theta^2F_M
\end{equation}
The scalar component $M$ may remain ultralight if protected by an approximate shift symmetry, sequestering, no scale structure, nonrenormalization effects, large volume suppression or exponentially small nonperturbative stabilization \cite{hs1cacciapaglia2024hidden,hs2feldman2010multicomponent,hs3blinov2012dark,hs4andreas2013dark,hs5braun2013supersymmetric,hs6strassler2006possible,hs7acharya2016lightest,hs8dine2004visible,hs9murayama2008more,hs10chan2012lhc,dec5arvanitaki2010string}. In a generic $N=1$ supergravity theory \cite{susy1fayet1977supersymmetry,susy2martin2010supersymmetry,susy3feng2013naturalness,susy4sohnius1985introducing,susy5witten1982constraints,dec3kane2015cosmological}, its mass is determined by the curvature of the scalar potential
\begin{equation}
m_\varphi^2 = \frac{1}{K_{M\bar M}}
\frac{\partial^2 V_{\rm SUGRA}}{\partial M\partial \bar M} \bigg|_{M=M_0}
\end{equation}
where we have
\begin{equation}
V_{\rm SUGRA} = e^{K/M_{\rm Pl}^2}
\left(
K^{i\bar j}D_iW D_{\bar j}\bar W
-
\frac{3|W|^2}{M_{\rm Pl}^2}
\right)
+ V_D
\end{equation}
and
\begin{equation}
D_iW = \partial_iW+\frac{\partial_iK}{M_{\rm Pl}^2}W
\end{equation}
Thus MAGIS/AION-type experiments probe precisely those supersymmetric sectors in which $m_\varphi$ lies in the ultralight window accessible to coherent phase searches. Furthermore, the effective coupling $d_{\rm eff}$ is not a phenomenological constant once a supersymmetric hidden sector is specified. It is determined by the dependence of visible sector parameters on the modulus direction \cite{dec1damour2010equivalence,dec4acharya2008nonthermal} as for the gauge sector one has \cite{dec2kaplan2000couplings}
\begin{equation}
{\cal L}
\supset
\int d^2\theta\,
\frac{1}{4}f_a({\cal M})W_a^\alpha W^a_\alpha
+
{\rm h.c.}
\end{equation}
with
\begin{equation}
\frac{1}{g_a^2} = {\rm Re}\,f_a(M)
\end{equation}
A fluctuation of the canonically normalized modulus $\varphi$ then gives
\begin{equation}
\frac{\delta\alpha}{\alpha}
= d_e\kappa\varphi
\end{equation}
where \begin{equation}
d_e = - \frac{1}{\kappa\sqrt{K_{M\bar M}}}
\frac{\partial\ln{\rm Re}f_{\rm EM}}{\partial M}
\end{equation}
Similarly, the electron mass depends on the modulus through the Kähler metrics, the Yukawa coupling and the Higgs sector and with
\begin{equation}
K \supset
Z_L(M,\bar M)L^\dagger L
+
Z_E(M,\bar M)E^\dagger E
+
Z_{H_d}(M,\bar M)H_d^\dagger H_d
\end{equation}

and

\begin{equation}
W
\supset
y_e(M)H_dLE
\end{equation}

the physical electron mass is schematically

\begin{equation}
m_e
=
\frac{e^{K/2M_{\rm Pl}^2}y_e(M)v_d(M)}
{\sqrt{Z_LZ_EZ_{H_d}}}
\end{equation}

so that

\begin{equation}
\frac{\delta m_e}{m_e}
=
d_{m_e}\kappa\varphi
\end{equation}

with

\begin{equation}
d_{m_e}
=
\frac{1}{\kappa\sqrt{K_{M\bar M}}}
\frac{\partial}{\partial M}
\ln
\left[
\frac{e^{K/2M_{\rm Pl}^2}y_e(M)v_d(M)}
{\sqrt{Z_LZ_EZ_{H_d}}}
\right]
\end{equation}

The QCD scale may also vary through the modulus dependence of the strong gauge kinetic function, since \begin{equation}
\Lambda_{\rm QCD}
=
\mu
\exp
\left[
-\frac{8\pi^2}{b_3g_3^2(\mu)}
\right]
\end{equation}
and $g_3^{-2}={\rm Re}f_3(M)$, so one obtains

\begin{equation}
\frac{\delta\Lambda_{\rm QCD}}{\Lambda_{\rm QCD}}
=
d_g\kappa\varphi
\end{equation}
where
\begin{equation}
d_g
=
-
\frac{8\pi^2}{b_3}
\frac{1}{\kappa\sqrt{K_{M\bar M}}}
\frac{\partial{\rm Re}f_3}{\partial M}
\end{equation}
The atomic transition frequency responds to these variations through sensitivity coefficients, as
\begin{equation}
\frac{\delta\omega_A}{\omega_A}
=
K_\alpha\frac{\delta\alpha}{\alpha}
+
K_{m_e}\frac{\delta m_e}{m_e}
+
K_g\frac{\delta\Lambda_{\rm QCD}}{\Lambda_{\rm QCD}}
+
K_q\frac{\delta m_q}{m_q}
+
K_H\frac{\delta v}{v}
+\cdots
\end{equation}
Hence for a supersymmetric modulus or dilaton, the experimentally measured effective coupling is
\begin{equation}
d_{\rm eff}^{\rm SUSY}
=
K_\alpha d_e
+
K_{m_e}d_{m_e}
+
K_gd_g
+
K_qd_q
+
K_Hd_H
+\cdots
\end{equation}
or, more explicitly
\begin{equation}
\begin{aligned}
d_{\rm eff}^{\rm SUSY}
=
&
-
K_\alpha
\frac{1}{\kappa\sqrt{K_{M\bar M}}}
\frac{\partial\ln{\rm Re}f_{\rm EM}}{\partial M}
\\
&
+
K_{m_e}
\frac{1}{\kappa\sqrt{K_{M\bar M}}}
\frac{\partial}{\partial M}
\ln
\left[
\frac{e^{K/2M_{\rm Pl}^2}y_e(M)v_d(M)}
{\sqrt{Z_LZ_EZ_{H_d}}}
\right]
\\
&
-
K_g
\frac{8\pi^2}{b_3}
\frac{1}{\kappa\sqrt{K_{M\bar M}}}
\frac{\partial{\rm Re}f_3}{\partial M}
+
K_qd_q
+
K_Hd_H
+\cdots
\end{aligned}
\end{equation}
This is the central relation between atom interferometry and supersymmetric hidden sectors, as the interferometer constrains $d_{\rm eff}^{\rm SUSY}$, while the microscopic theory determines $d_{\rm eff}^{\rm SUSY}$ from derivatives of gauge kinetic functions, Kähler metrics, Yukawa couplings, Higgs sector parameters and the strong coupling scale along the light modulus direction.
\\
\\
The modulus abundance enters through $F_{\rm DM}$. If the scalar begins coherent oscillations when \cite{dec4acharya2008nonthermal}
\begin{equation}
3H(t_{\rm osc})
\simeq
m_\varphi
\end{equation}
then its initial energy density is
\begin{equation}
\rho_\varphi(t_{\rm osc})
\simeq
\frac{1}{2}m_\varphi^2\varphi_i^2
\end{equation}
and subsequently redshifts as nonrelativistic matter, so $F_{\rm DM}$ is determined by the initial displacement, the scalar mass, the cosmological expansion history and possible dilution or decay channels. The atom interferometer limit hence applies to the combination $d_{\rm eff}^{\rm SUSY}\sqrt{F_{\rm DM}}$, rather than to $d_{\rm eff}^{\rm SUSY}$ alone. Putting these ingredients together, the supersymmetric hidden sector phase signal is given by
\begin{equation}
\Delta\Phi_{\rm SUSY}(t)
=
\omega_A T_{\rm eff}
d_{\rm eff}^{\rm SUSY}
\kappa
\frac{\sqrt{2F_{\rm DM}\rho_{\rm DM}}}{m_\varphi}
\cos(m_\varphi t+\beta)
\end{equation}
A null result gives the fully symbolic bound

\begin{equation}
\left|d_{\rm eff}^{\rm SUSY}\right|
<
\frac{A_{\rm lim}}{\omega_A T_{\rm eff}}
\frac{m_\varphi}{\kappa\sqrt{2F_{\rm DM}\rho_{\rm DM}}}
\end{equation}
Equivalently, in terms of the underlying modulus dependence we have
\begin{multline}
\left|
K_\alpha d_e
+
K_{m_e}d_{m_e}
+
K_gd_g
+
K_qd_q
+
K_Hd_H
+\cdots
\right|
\\ < 
\frac{A_{\rm lim}}{\omega_A T_{\rm eff}}
\frac{m_\varphi}{\kappa\sqrt{2F_{\rm DM}\rho_{\rm DM}}}
\end{multline}
explictly, we have
\begin{multline}
\Bigg|
-
K_\alpha
\frac{1}{\kappa\sqrt{K_{M\bar M}}}
\frac{\partial\ln{\rm Re}f_{\rm EM}}{\partial M}
+ \\
K_{m_e}
\frac{1}{\kappa\sqrt{K_{M\bar M}}}
\frac{\partial}{\partial M}
\ln
\left[
\frac{e^{K/2M_{\rm Pl}^2}y_e(M)v_d(M)}
{\sqrt{Z_LZ_EZ_{H_d}}}
\right]
- \\
K_g
\frac{8\pi^2}{b_3}
\frac{1}{\kappa\sqrt{K_{M\bar M}}}
\frac{\partial{\rm Re}f_3}{\partial M}
+
K_qd_q
+
K_Hd_H
+\cdots
\Bigg|
\\
<
\frac{A_{\rm lim}}{\omega_A T_{\rm eff}}
\frac{m_\varphi}{\kappa\sqrt{2F_{\rm DM}\rho_{\rm DM}}}
\end{multline}
This is our main constraint equation and here it is deliberately written in symbolic form because the final experimental limit must be obtained from the measured phase noise spectrum and the exact MAGIS/AION response function. The equation regardless shows us that a long-baseline atom interferometer constrains the geometry of the hidden sector itself, with the visible sector derivatives of $f_a$, $Z_i$, $y_i$, $v_d$ and $\Lambda_{\rm QCD}$ along an ultralight supersymmetric direction.
\\

For illustration only, one may insert representative numbers, where we take $\rho_{\rm DM}\simeq0.4\,{\rm GeV/cm^3}$, which gives us

\begin{equation}
\kappa\varphi_0
\simeq
7.2\times10^{-16}
\left(
\frac{10^{-15}\,{\rm eV}}{m_\varphi}
\right)
\sqrt{F_{\rm DM}}
\end{equation}

For an optical transition with $\nu_A\simeq4.3\times10^{14}\,{\rm Hz}$, close to the strontium clock transition used in AION/MAGIS-type proposals \cite{aion1Badurina:2019hst,magis1abe2021matter}, one has $\omega_A\simeq2.7\times10^{15}\,{\rm s}^{-1}$ and with $T_{\rm eff}\simeq1\,{\rm s}$,
\begin{equation}
A_{\rm SUSY}
\simeq
1.9
\left(
\frac{10^{-15}\,{\rm eV}}{m_\varphi}
\right)
\sqrt{F_{\rm DM}}
d_{\rm eff}^{\rm SUSY}
\end{equation}
If a future atom interferometric search reaches an effective phase amplitude limit $A_{\rm lim}\sim10^{-8}\,{\rm rad}$ \cite{aionsenseBadurina:2021lwr,aionsense2Badurina:2022ngn,aionsense3Badurina:2023wpk,aionsense4Badurina:2021rgt} at a given mass, the corresponding illustrative coupling reach is
\begin{equation} \label{defconst}
\left|d_{\rm eff}^{\rm SUSY}\right|
\lesssim
5.3\times10^{-9}
\left(
\frac{A_{\rm lim}}{10^{-8}\,{\rm rad}}
\right)
\left(
\frac{m_\varphi}{10^{-15}\,{\rm eV}}
\right)
\frac{1}{\sqrt{F_{\rm DM}}}
\left(
\frac{1\,{\rm s}}{T_{\rm eff}}
\right)
\end{equation}
Thus, for $F_{\rm DM}=1$ and $T_{\rm eff}=1\,{\rm s}$,
\begin{equation}
m_\varphi=10^{-17}\,{\rm eV}
\quad\Rightarrow\quad
\left|d_{\rm eff}^{\rm SUSY}\right|
\lesssim
5\times10^{-11}
\end{equation}
\begin{equation}
m_\varphi=10^{-15}\,{\rm eV}
\quad\Rightarrow\quad
\left|d_{\rm eff}^{\rm SUSY}\right|
\lesssim
5\times10^{-9}
\end{equation}
\begin{equation}
m_\varphi=10^{-12}\,{\rm eV}
\quad\Rightarrow\quad
\left|d_{\rm eff}^{\rm SUSY}\right|
\lesssim
5\times10^{-6}
\end{equation}
These numbers should not be read as final MAGIS or AION bounds. They are meant only as benchmark projections, useful for showing the parametric reach of the method. In an actual analysis, $A_{\rm lim}$ would have to be replaced by the experimentally measured, frequency-dependent phase limit, with the transfer function, environmental backgrounds, gravity gradient noise and possible network correlations all included.
\begin{figure*}
\centering
\includegraphics[width=0.8\textwidth]{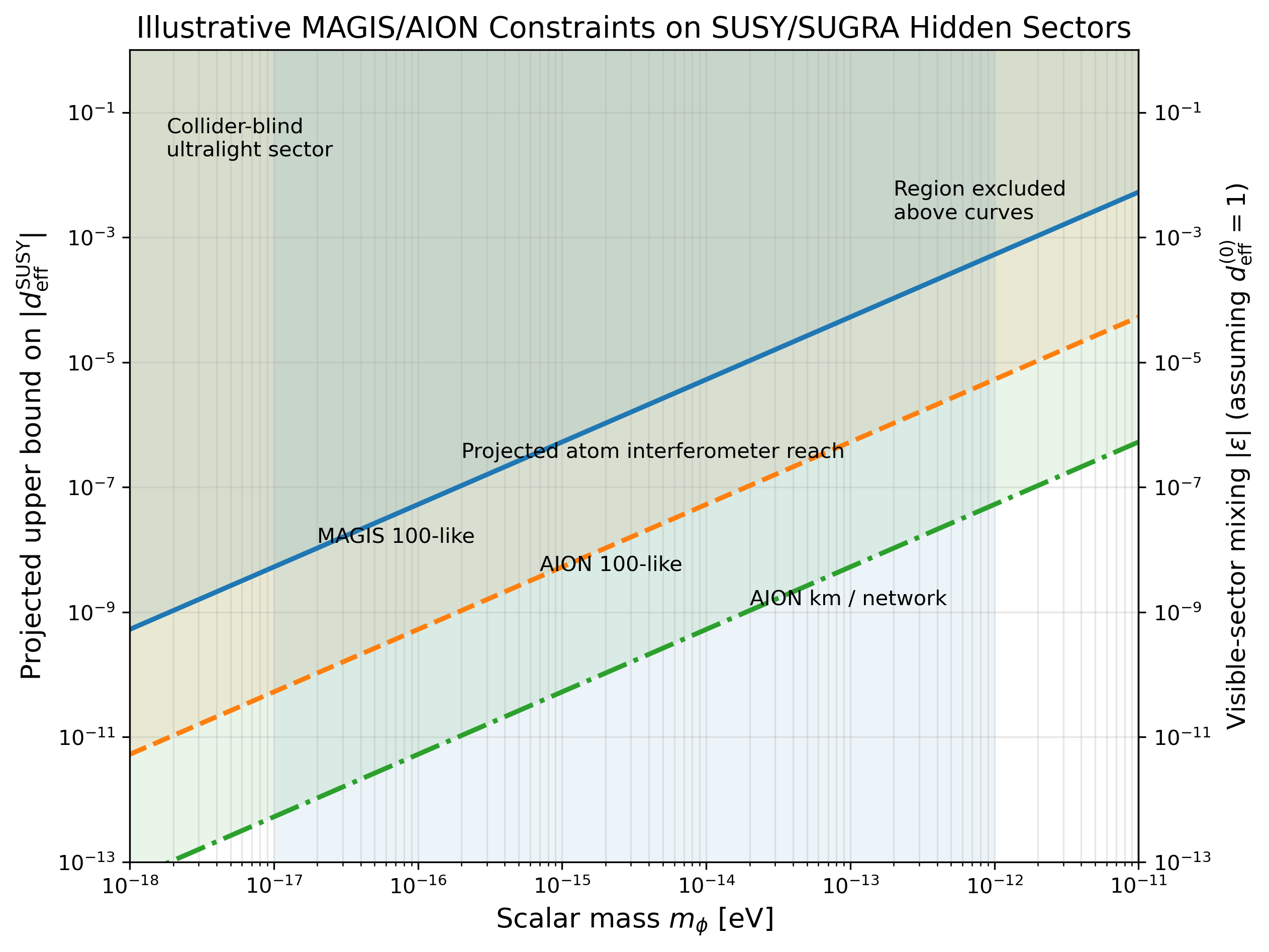}
\caption{Illustrative projected constraints on supersymmetric hidden sector scalars from MAGIS/AION-type atom interferometers in the $(m_\phi,|d_{\rm eff}^{\rm SUSY}|)$ plane, in accordance to Eq. \eqref{defconst}. The curves assume $F_{\rm DM}=1$ , $T_{\rm eff}=1 \, {\rm s}$, $d_{\rm eff}^{(0)}=1$ for the right axis and benchmark phase amplitude limits rather than final experimental exclusion curves.}
\label{atomsusy}
\end{figure*}
Fig. \ref{atomsusy} summarizes the main result in a simple way, as there we show how a null search for oscillatory phase signals in a long-baseline atom interferometer can be translated into a bound on the effective supersymmetric hidden sector coupling $d_{\rm eff}^{\rm SUSY}$. The approximate scaling is $|d_{\rm eff}^{\rm SUSY}|\propto m_\phi A_{\rm lim}/\sqrt{F_{\rm DM}}$, so for a fixed phase sensitivity the reach is stronger for lighter coherent fields. The right axis gives the corresponding visible sector mixing angle $\epsilon$, assuming an unsuppressed modulus coupling $d_{\rm eff}^{(0)}\sim 1$. In this interpretation, MAGIS/AION type instruments could probe very small visible admixtures of ultralight fields that otherwise remain hidden. We stress again that these curves are only benchmark projections. A final experimental study would need to use the full frequency dependent response and the actual measured phase noise spectrum.
\\

A simple way to interpret the result is to suppose that the light scalar is mostly hidden, and talks to the visible sector only through a small mixing angle $\epsilon$. In that case one may write
\begin{equation}
\varphi_{\rm light}=
\cos\epsilon \, \varphi_{\rm hidden}
+
\sin\epsilon \, \varphi_{\rm vis}
\end{equation}
If the unsuppressed visible modulus has $d_{\rm eff}^{(0)}\sim{\cal O}(1)$, then
\begin{equation}
d_{\rm eff}^{\rm SUSY}
\simeq
\epsilon d_{\rm eff}^{(0)}
\end{equation}
and the atom interferometric constraint becomes
\begin{equation}
\epsilon
\lesssim
\frac{1}{d_{\rm eff}^{(0)}}
\frac{A_{\rm lim}}{\omega_A T_{\rm eff}}
\frac{m_\varphi}{\kappa\sqrt{2F_{\rm DM}\rho_{\rm DM}}}
\end{equation}
illustratively, this is
\begin{equation}
\epsilon
\lesssim
5.3\times10^{-9}
\left(
\frac{A_{\rm lim}}{10^{-8} \,{\rm rad}}
\right)
\left(
\frac{m_\varphi}{10^{-15} \,{\rm eV}}
\right)
\frac{1}{d_{\rm eff}^{(0)}\sqrt{F_{\rm DM}}}
\left(
\frac{1 \,{\rm s}}{T_{\rm eff}}
\right)
\end{equation}
Thus long-baseline atom interferometers can be sensitive to extremely small visible sector admixtures of ultralight supersymmetric hidden fields. This gives a near Earth way of testing sequestered moduli, dilatons and hidden scalars whose masses are far below collider thresholds and whose direct production would not be realistic.
\\
\\
The main point of this result is that it gives a non-collider way of looking for supersymmetric hidden sectors and string-motivated moduli physics. MAGIS and AION would not be searching for superpartners through high energy production, as collider experiments do, but for coherent low energy remnants of such theories through atomic phase shifts. A null result would constrain the derivatives of gauge kinetic functions, Kähler metrics, Yukawa couplings, Higgs sector parameters and QCD scale dependence along ultralight hidden sector directions. On the other hand, a detection would appear as a narrow oscillatory phase line. Its frequency would determine $m_\varphi$, its amplitude would determine the combination $d_{\rm eff}^{\rm SUSY}\sqrt{F_{\rm DM}}$, and its response in different atomic transitions could help separate photon, electron, QCD or Higgs sector couplings. In this sense, atom interferometry makes supersymmetric moduli and dilatons less of a purely ultraviolet model building ingredient, and turns them into something that can be tested through near Earth precision measurements. These constraints would also be complementary to collider searches, equivalence principle tests and astrophysical probes. Thus, this setup gives a useful precision window onto hidden sectors which may remain present even if visible sector supersymmetry is not directly accessible. \\

What models are actually being tested here? The most direct targets are not generic supersymmetric models, but SUSY or string motivated constructions in which an ultralight scalar direction controls some visible sector parameters. Examples include KKLT type and large volume moduli stabilization scenarios, no scale or sequestered supergravity models, dilaton-like theories with varying gauge couplings, supersymmetric hidden valleys, supersymmetric dark-sector models and axiverse-inspired constructions where modulus partners can remain light \cite{dec1damour2010equivalence,dec2kaplan2000couplings,dec3kane2015cosmological,dec4acharya2008nonthermal,dec5arvanitaki2010string,kkltKachru:2003aw,lv1Balasubramanian:2005zx,lv2Damour:1994zq}. The constraint discussed here clearly cannot rule out supersymmetry or string theory as full frameworks. It also says nothing about models where all moduli are heavy, cosmologically irrelevant, or completely sequestered from the Standard Model. What it can test is the subset of low energy SUSY/SUGRA Lagrangians in which an ultralight modulus, dilaton or hidden scalar lies in the MAGIS/AION mass range, has a non-negligible abundance, and couples to the visible sector strongly enough to produce an observable atomic phase modulation. This is what makes the probe interesting, because it constrains the infrared leakage of hidden sector geometry into atomic observables, namely the dependence of $f_a$, $Z_i$, $y_i$, $v_d$, and $\Lambda_{\rm QCD}$ on light supersymmetric directions, rather than the production of visible superpartners or the high energy spectrum itself. A null result would place bounds on sequestering and coupling geometry for ultralight hidden sectors, while a positive signal would point to a coherent hidden relic whose coupling pattern could distinguish electromagnetic, fermion mass, QCD and Higgs sector origins. \\

\textbf{Acknowledgements:}
The work of OT is supported in part by the Vanderbilt Discovery Doctoral Fellowship. The author thanks Robert J. Scherrer, Abraham Loeb and Alfredo Gurrola for helpful discussions on the topics of this work.  
\bibliography{apssamp}

\end{document}